\documentclass{article}

\usepackage{PRIMEarxiv}

\usepackage[utf8]{inputenc} 
\usepackage[T1]{fontenc}    
\usepackage{hyperref}       
\usepackage{url}            
\usepackage{booktabs}       
\usepackage{amsfonts}       
\usepackage{nicefrac}       
\usepackage{microtype}      
\usepackage{lipsum}
\usepackage{fancyhdr}       
\usepackage{graphicx}       
\graphicspath{{media/}}     

\title{Mixed vine copula flows for flexible modelling of neural dependencies
\thanks{\textit{\underline{Corresponding author}}: 
\textbf{Lazaros Mitskopoulos, L.Mitskopoulos@sms.ed.ac.uk}} 
}

\author{
  Lazaros Mitskopoulos \\
  School of Informatics \\
  University of
Edinburgh \\
  Edinburgh\\
  \texttt{L.Mitskopoulos@sms.ed.ac.uk} \\
   \And
  Theoklitos Amvrosiadis \\
  Centre for Discovery Brain Sciences \\
  University of Edinburgh \\
  Edinburgh\\
  \texttt{t.amvrosiadis@ed.ac.uk} \\
  \And
  Arno Onken \\
  School of Informatics \\
  University of Edinburgh \\
  Edinburgh \\
  \texttt{aonken@inf.ed.ac.uk} \\
}

\DeclareUnicodeCharacter{2212}{-}
\begin{document}
\maketitle

\begin{abstract}
Recordings of complex neural population responses provide a unique opportunity for advancing our understanding of neural information processing at multiple scales and improving performance of brain computer interfaces. However, most existing analytical techniques fall short of capturing the complexity of interactions within the concerted population activity. Vine copula-based approaches have shown to be successful at addressing complex high-order dependencies within the population, disentangled from the single-neuron statistics. However, most applications have focused on parametric copulas which bear the risk of misspecifying dependence structures. In order to avoid this risk, we adopted a fully non-parametric approach for the single-neuron margins and copulas by using Neural Spline Flows (NSF). We validated the NSF framework on simulated data of continuous and discrete type with various forms of dependency structures and with different dimensionality. Overall, NSFs performed similarly to existing non-parametric estimators, while allowing for considerably faster and more flexible sampling which also enables faster Monte Carlo estimation of copula entropy. Moreover, our framework was able to capture low and higher order heavy tail dependencies in neuronal responses recorded in the mouse primary visual cortex during a visual learning task while the animal was navigating a virtual reality environment. These findings highlight an often ignored aspect of complexity in coordinated neuronal activity which can be important for understanding and deciphering collective neural dynamics for neurotechnological applications. 
\end{abstract}


\section{Introduction}
Coordinated information processing by neuronal circuits in the brain is the basis of perception and action. Neuronal ensembles encode sensory and behavior-related features in sequences of spiking activity which can exhibit rich dynamics at various temporal scales \cite{gao2015simplicity}. Acquiring an understanding of how multivariate interactions in neural populations shape and affect information transmission is not only important for neural coding theory but will also inform methodological frameworks for clinically translatable technologies such as Brain Computer Interfaces (BCIs). Both of these research programs have enjoyed a surge of activity as a result of recent advances in imaging technologies \cite{chen2012lotos} and high-yield electrophysiology both for human \cite{mcfarland2017eeg} and animal studies \cite{jun2017fully}. BCIs can mediate neural signal transduction for moving prosthetic limbs or external robotic devices in paralyzed patients or they can aid communication with patients suffering from locked-in syndrome \cite{chaudhary2016brain}. Therefore, successful clinical use relies on accurate reading and relaying of information content transmitted via population spiking responses. Doing so can be quite challenging from a data analytic perspective as moderate to high-dimensional brain activity can be considerably complex, exhibiting non trivial multivariate neural dependencies \cite{hurwitz2021building}. Moreover, the resulting behavioral output variables (e.g. limb movement) might display vastly different statistics to neural variables like spike trains or event-related potentials (ERPs). These challenges highlight the importance of developing novel analytical tools that can handle the complexity within neural population activity and its relation to behavior. Such tools should also have broad applicability over different types of data (e.g. spike counts, local field potentials, EPRs). 

The present need for novel methods stems from the fact that the majority of past work on neural dependencies has focused on pairwise shared response variability between neurons, also known as noise correlations \cite{zohary1994correlated, brown2004multiple,moreno2014information,kohn2016correlations}. Neural responses are known to exhibit considerable variability even when repeatedly presented with the same stimulus, but this might be part of collective dynamical patterns of activity in a neural population. The typical assumption in this line of research is that the noise in neural responses is Gaussian, and thus, firing rates are modelled with multivariate normal distributions where a certain covariance matrix specifies all pairwise linear correlations \cite{averbeck2006neural,ecker2010decorrelated}. While this approach may provide a reasonable approximation for correlations in coarse time-scales, its validity can be disputed for spike-counts in finer time-scales. First of all, real spike counts are characterized by discrete distributions and they exhibit a positive skew instead of a symmetric shape \cite{onken2009analyzing}. Also, spike data do not usually display elliptical dependence as in the normal distribution but tend to be heavy tailed \cite{kudryashova2022parametric}, which geometrically translates to having probability mass concentrated on one of the corners. Finally, although the multivariate normal distribution is characterized by only second-order correlations, recent studies have indicated that higher order correlations are substantially present in local cortical populations and have a significant effect on the informational content of encoding \cite{pillow2008spatio,ohiorhenuan2010sparse,yu2011higher,shimazaki2012state,montangie2017higher} as well as on the performance of decoding models \cite{michel2006costs}.
Therefore, dissecting the structure of multivariate neural interactions is important to the study of neural coding and clinical applications such as BCIs that rely on accurate deciphering of how neural activity translates to action. This calls for an alternative approach that goes beyond pairwise correlations. 

A statistical tool which is suited for the study of multivariate dependencies is that of copulas. Copula-based approaches have enjoyed wide usage in economics for modelling risk in investment portfolios \cite{jaworski2012copulae}, but have received limited attention in neuroscience. Intuitively, copulas describe the dependence structures between random variables, and in conjunction with models of the individual variables, they can form a joint statistical model for multivariate observations. When observations come from continuous variables their associated copulas are unique, independent from marginal distributions and invariant to strictly monotone transformations \cite{faugeras2017inference}. However, data in neuroscience research are often discrete (e.g. spike counts) or they may contain interactions between discrete and continuous variables with vastly different statistics (e.g. spikes with local field potentials or behavioral measurements such as running speed or pupil dilation). Despite the indeterminacy of copulas in these cases, they are still valid constructions for discrete objects and mixed interactions and one can apply additional probabilistic tools to obtain consistent discrete copulas \cite{genest2007primer}. Previous work has successfully applied copula-based methods to discrete or mixed settings \cite{song2009joint,de2011copula,smith2012estimation, panagiotelis2012pair,onken2016mixed} using copulas from parametric families that assume a certain type of interaction. Although parametric models are a powerful tool for inference, their application can bear a risk of misspecification by imposing rather rigid and limiting assumptions on the types of dependencies to be encountered within a dataset with heterogeneous variables or multiscale dynamical processes. This risk is especially amplified for dependencies between more than two variables as available multivariate copulas are quite limited in number and assume a particular type of dependence structure for all variables which can ignore potentially rich interaction patterns. As the set of commonly used bivariate parametric copulas is much larger, a common alternative is to decompose multivariate dependencies into a cascade of bivariate copulas organized into hierarchical tree structures called vines or pair copula constructions \cite{aas2009pair}. Nodes of the vines correspond to conditional distributions and edges correspond to copulas that describe their interaction. This formulation allows for a flexible incorporation of various dependence structures in a joint model. Previous studies that employed vine copulas in mixed settings used parametric models \cite{song2009joint,de2011copula,smith2012estimation, panagiotelis2012pair,onken2016mixed} For the present study, given the aforementioned intricacies of neuronal spiking statistics, we aim to explore the potential of non-parametric methods as a more flexible alternative for estimating discrete and continuous vine copulas. Existing non-parametric approaches have focused on kernel-based methods \cite{racine2015mixed,geenens2017probit} or jittering and continuous convolutions with specified noise models to obtain pseudo-continuous data \cite{schallhorn2017d,nagler2017nonparametric}.

Another model-free method that shows promise is that of normalizing flows, a class of generative models for density estimation that allow for flexible sampling \cite{rezende2015variational,papamakarios2021normalizing}. Recently, some authors have attempted to employ normalizing flows for non-parametric modeling of copulas using simulated and standard benchmark datasets \cite{wiese2019copula,kamthe2021copula}. An application of these models to recordings from neural populations is still missing and has the potential to shed light on the structure of coordinated neural activity, thereby potentially improving BCIs that take this structure into account.

In this study, we aimed to conduct a thorough investigation into flow-based estimation of vine copulas with continuous and discrete artificial data in various settings with different but known dependence structures and number of variables. Furthermore, we sought to demonstrate the potential of this framework to elucidate interaction patterns within neural recordings that contain heavy tails and extend beyond bivariate dependencies. For this reason, we chose to investigate neural responses in the mouse V1 while the animal is navigating in a virtual reality environment. Studying neural interfaces in rodents has been important for pre-clinical testing of BCIs to probe potential limitations that can inform applications in humans \cite{widge2014pre,bridges2018rodent}. The test case we chose serves as a proof-of-concept study but it can also provide meaningful insights on how spatial navigation related cues and/or behavioral variables modulate visual activity, which can inform future clinical research on BCIs.

\section{Materials and Methods}
\subsection{Copulas}
Multivariate neuronal interactions can be described probabilistically by means of copulas, which embody how spiking statistics of individual neurons, i.e. the marginal distributions, are entangled in intricate ways to produce the observed joint population activity. The central theoretical foundation for copula-based approaches is Sklar’s theorem \cite{sklar1959fonctions}. It states that every multivariate cumulative distribution function (CDF) $F_x$ can be decomposed into its margins, in this case the single-neuron distributions $F_1,...F_d$, and a copula $C$  (Figure 1A) such that:
\begin{equation}
\label{eqn:sklar1}
 F_x (x_1 , \dots , x_d) = C(F_1(x_1), \dots , F_d(x_d))  
\end{equation}

\begin{figure}
    \centering
    \includegraphics[scale=0.25]{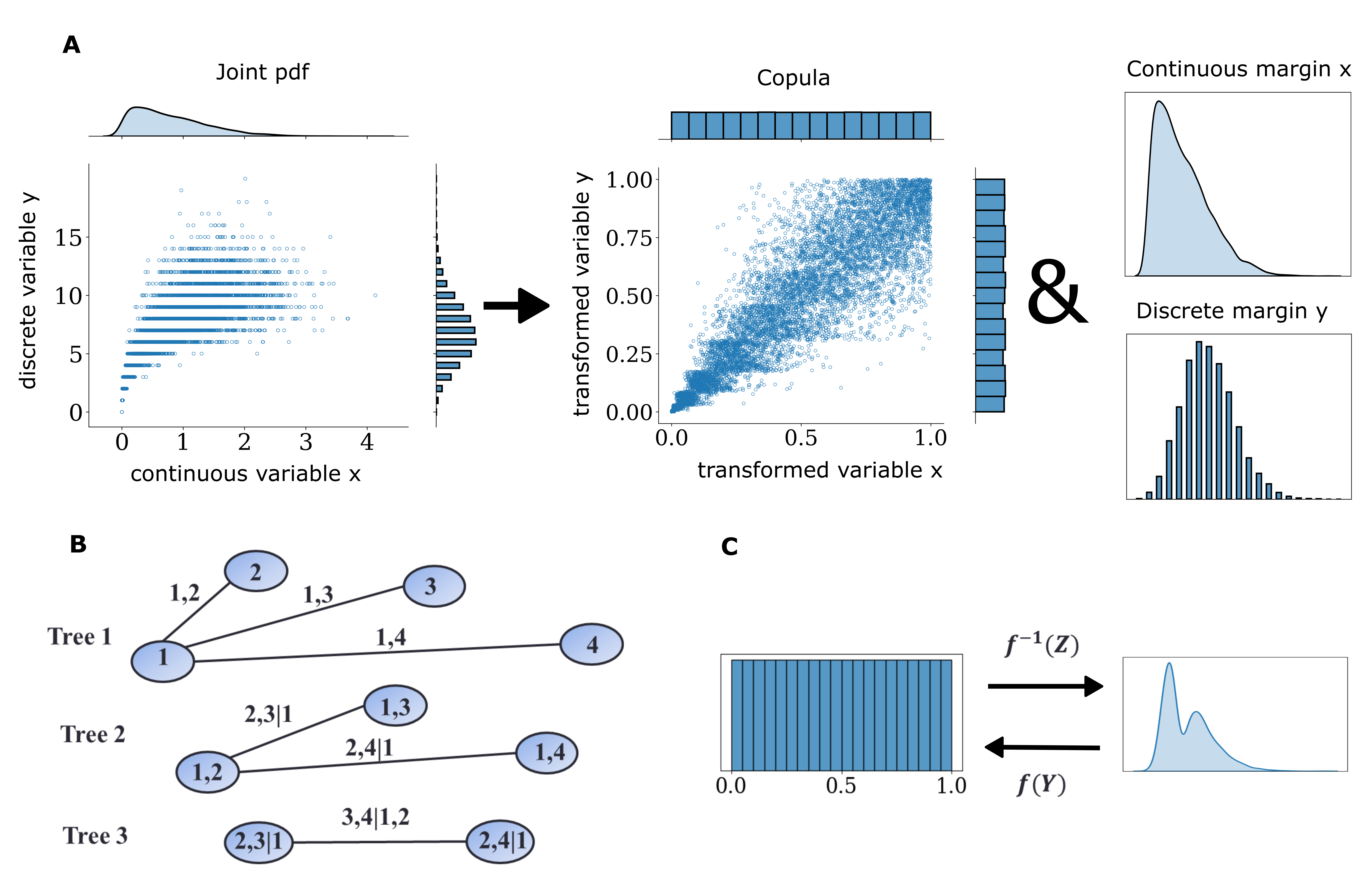}
    \caption{Mixed vine copula flows. \textbf{A}. Samples from mixed variables of any joint probability density function (pdf) can be decomposed into their margins and a copula. Copulas are extracted by transforming continuous and discrete samples to uniform through the probability integral transform. \textbf{B}. Graphical illustration of a C-vine for 4 variables. Nodes and edges of the first tree denote the variables and bivariate dependencies respectively. Edges of subsequent trees denote dependencies that condition on one or more variables. \textbf{C}. Illustration of forward and inverse normalizing flow transformation \textit{f} of base distribution \textit{Z} and target distribution \textit{Y}.}
    \label{fig:cop_vines_flows}
\end{figure}

 Copulas are also multivariate CDFs with support on the unit hypercube and uniform margins and their shape describes the dependence structure between random variables in a general sense which goes beyond linear or rank correlations \cite{faugeras2017inference}. Following Sklar's theorem, it is possible to obtain copulas from joint distributions using:

\begin{equation}
\label{eqn:sklar2}
C(u_1, \dots, u_d) = F_x(F^{-1}_1(u_1), \dots, F^{-1}_d(u_d)),
\end{equation}

 Conversely, it is also possible to construct proper joint distributions by entangling margins with copulas. These operations rely on the probability transform $F$ and its generalized inverse, the quantile transform $F^{-1}$. The probability transform maps samples to the unit interval: $: F(X) \rightarrow{U \sim{U_{[0,1]}}}$, where $U_{[0,1]}$ denotes the uniform distribution on the unit interval. Since copulas for discrete data depend on margins and are not unique \cite{genest2007primer}, additional tools are required to obtain consistent mapping to copula space. We employed the distributional transform: 
 
 \begin{equation}
\label{eqn:dist_trans}
G(X,V)= F_{x-}(x) + V(F_x(x)-F_{x-}(x))
 \end{equation}

where $F_{x-}(x)=Pr(X < x) $ as opposed to the regular expression for the CDF, $F_x(x)=Pr(X <= x) $, and $V$ is a random variable uniformly distributed on [0,1] independent of $X$. This extension to the probability transformation effectively converts a discrete copula to a pseudo-continuous variable by adding uniform jitter in between discontinuous intervals in the support of discrete variables and makes it possible to use non-parametric estimators designed for continuous data. An example with continuous and discrete observations is illustrated in Figure 1A. This case of a mixed interaction is described by a Clayton copula which displays an asymmetric heavy tail. The empirical copula can be discovered by subjecting the variables to the probability transform with an added distributional transform for the discrete one. This operation dissects the dependence information that is embedded in the joint probabilities of these two variables.

\subsection{Pair copula constructions}
 The curse of dimensionality encountered in large datasets can pose considerable challenges for copula modeling. Pair copula constructions \cite{bedford2002vines} offer a flexible way of scaling copula models by factorising the multivariate distribution into a cascade of bivariate conditional distributions that can be described by bivariate copulas. The latter can be modelled parametrically, as previous studies have already done \cite{onken2016mixed,panagiotelis2012pair} or with non-parametric tools, which is the present study's approach.
 
 The space of possible factorizations is prohibitively large so vine copula structures, a special type of pair copula constructions can be employed to facilitate inference and sampling \cite{aas2009pair}. They can be represented as hierarchical sets of trees which account for a specific graph of multivariate interactions among elements of the distributions and assume conditional independence for the rest. In the present study, we focused on the canonical vine or C-Vine (Figure 1B) in which each tree in the hierarchy has a node that serves as a connection hub to all other nodes. The C-Vine decomposes the joint distribution $f$ into a product of its margins and conditional copulas $c$.
 
 \begin{equation}
     f_X(x_1,... , x_d) = \prod_{k=1}^{d}f(x_k) \prod_{j=1}^{d-1} \prod_{i=1}^{d-j} c_{j,i+j|1, ... , j−1}(F(x_j |x_1, ... , x_{j-1}), F(x_{i+j} |x_1, ... , x_{j-1}))
 \end{equation}
 
where $c_{i,j|A}$ denotes the pair copula between elements $i$ and $j$ given the elements in the set $A$", which is empty in the surface tree but it increases in number of elements with deeper trees. 

\subsection{Copula Flows}
We modeled the margin and copula densities non-parametrically using a specific class of normalizing flows, that is called Rational-Quadratic Neural Spline Flows (NSF) \cite{durkan2019neural}. In general, normalizing flows are a class of generative models that construct arbitrary probability densities using smooth and invertible transformations to and from simple probability distributions \cite{rezende2015variational} (Figure 1C). In essence, this is an application of the change of variables formula:
 \begin{equation}
\label{eqn:flow_main_eqn}
p_x(x)= p_u(T^{-1}(x))\det\left|\frac{\partial T^{-1}(x)}{\partial x}\right|,
 \end{equation}

where $p_x(x)$ is the density of the observations and $p_u$ is the base density of random variable $U=T^{-1}(X)$, which is a known and convenient distribution such as the normal or the uniform distribution. The transformation $T$ is usually a composition of invertible transformations that can be perceived and implemented as an artificial neural network with a certain number of layers and hidden units. Its parameters have to be learned through training in order to achieve an invertible mapping between the two distributions, while scaling appropriately by the determinant of the Jacobian matrix which keeps track of volume changes induced by $T$.

Since our main goal was to apply normalizing flows on a copula-based framework for neural dependencies, it was natural to choose the uniform distribution on [0,1] as a base distribution, so that backward and forward flow transformations for the margins approximate the probability transform and its inverse, the quantile transform respectively, so as to map observations to copula space and back. Furthermore, a uniform base distribution for copula flows can be leveraged to generate new (simulated) observations via inverse sampling. Different types of normalizing flows exist in the literature which involve simple affine or more flexible non-affine transformations albeit with the cost of sacrificing invertibility in some cases \cite{papamakarios2021normalizing}. Our choice of employing NSF \cite{durkan2019neural} in this study for modelling both margin and copula densities was in virtue of the fact that they combine the flexibility of non-affine flows while maintaining easy invertibility by approximating a quasi-inverse with piecewise spline-based transformations around knot points of the mapping.

\subsection{Sequential Estimation and Model Selection}
 To fit the C-vine model with NSF to data, we applied the Inference for Margins procedure, which first fits the margins and then fits the copulas deriving from pairs of margins. For the first step, we fit each margin with NSF and then proceeded with the copulas of a particular canonical vine formulation \cite{aas2009pair}. Fitting NSF to bivariate copulas in the vine was conducted sequentially starting with the copulas of the surface layer of the tree. Subsequently, conditional marginals for the tree in the layer next in depth were estimated using the copulas of the previous layer via h-functions \cite{czado2019analyzing}. Then, conditional copulas were constructed by transforming the conditional marginals in the same layer according to (2). This procedure was followed until all copulas in the vine decomposition were estimated. For each copula, we used a random search procedure \cite{bergstra2012random} to determine the best performing NSF hyperparameter configuration on a validation set containing 10\% of the data. The hyperparameters that were tuned during training were the number of hidden layers and hidden units as well as the number of knots for the spline transforms. This sequential estimation and optimization scheme for NSF-based vine copulas was followed in both our analyses with artificial data as well as data from neuronal recordings.

\subsection{Other non-parametric estimators}
The other non-parametric estimators used for comparisons against NSF included four versions of Kernel Density Estimators (KDE), namely one with log-linear local likelihood estimation (tll1), one with log-quadratic local likelihood estimation (tll2), one with log-linear local likelihood estimation and nearest neighbour bandwidths (tll1nn) and one with log-quadratic local likelihood estimation and nearest neighbour bandwidths (tll2nn) \cite{nagler2017nonparametric}. Lastly, an estimator based on Bernstein polynomials \cite{sancetta2004bernstein} was also used for the comparisons. The implementations for all these five non-parametric estimators are from kdecopula package \cite{nagler2017nonparametric}.

\subsection{Artificial data}
The NSF framework for vine copulas was validated on artificial data with known dependency structures and was compared against other non-parametric estimators. We constructed several test cases where data was continuous or discrete, consisting of 4 (low dimensional) or 8 (higher dimensional) variables exhibiting weak or strong dependencies that were characterized by different copula families, namely either Clayton, Frank or Gaussian copulas. These three types of parametric copulas display different behavior in the tail regions \cite{nelsen2007introduction}. Clayton copulas have an asymmetric tail dependence whereby probability mass is concentrated in one corner of the copula space indicating a single heavy tail region (see example copula in Figure 1A). On the contrary, Frank copulas do not have heavy tails and probability mass is allocated uniformly and symmetrically along the correlation path. Gaussian copulas are also symmetric and without particularly heavy tails, but probability mass concentration in the tail regions is larger compared to Frank copulas. The strength of dependencies was determined by varying the $\theta$ parameter for Clayton ($\theta = 2$ for weak and $\theta = 5$ for strong dependence) and Frank copulas ($\theta = 3$ for weak and $\theta = 7$ for strong dependence) and the off-diagonal entries of the 2x2 correlation matrix for Gaussian copulas (0.4 for weak dependence and 0.8 for strong dependence). We constructed and drew simulated samples from all vines with the aforementioned specifications using the mixed vines package developed by Onken and Panzeri (2016). Training for all the estimators was conducted with 5000 simulated samples for each variable in the artificial data. The training procedure was repeated 10 times. Performance was measured with Kullback-Leibler (KL) divergences of 8000 copula samples generated by each of the estimators to an equal number of ground truth copula samples from the mixed vines package \cite{onken2016mixed}. To estimate the KL divergences from sample data, a k-nearest neighbour algorithm was used \cite{wang2009divergence}, which was implemented in \cite{nkl_hartland}. The resulting KL divergences from the 10 repetitions and the different copulas in each vine were aggregated to measure the overall performance of the framework in each test case. We statistically compared performances by all estimators via Kruskal-Wallis significance tests at level of significance equal to 0.05. Bonferonni correction was used for multiple comparisons. Moreover, we calculated copula entropies from the NSF copula densities via classical Monte Carlo estimation:
\begin{equation}
    h(c(u_1,u_2))=\mathbb{E}_c [-\log_2  c(u_1,u_2)] \approx -\frac{1}{k} \sum_{k=1}^K (\log_2  c(u^k_1,u^k_2)),
\end{equation}
where $h$ denotes the entropy and $\mathbb{E}_c$ denotes the expectation with respect to the copula $c$. The expectation is approximated by summing over a $K$ number of samples which needs to be sufficiently large ($K=8000$ in our study). Negative copula entropy provides an accurate estimate of mutual information between variables which does not depend on the marginal statistics of each variable \cite{sklar1959fonctions,jenison2004shape} and is a measure of the degree to which knowing one variable can reduce the uncertainty of the other one. All analyses including sampling and entropy calculations were conducted on an ASUS laptop with Intel(R) 4 Cores, i5-8300 CPU, 2.30 GHz.

\subsection{Experimental data}
In order to assess our framework's applicability to neuronal activity we used 
2-photon calcium imaging data of neuronal activity in the mouse primary visual cortex, that were was collected at the Rochefort lab (see \cite{henschke2020reward} for more details). Briefly, V1 layer 2/3 neurons labeled with the calcium indicator GCamP6s  were imaged while the animal was headfixed, freely running on a cylindrical treadmill and navigating through a virtual reality environment (160 cm). Mice were trained to lick at a specific location along the virtual corridor in order to receive a reward. In addition to neuronal activity, behavioral variables such as licking and running speed were monitored simultaneously. Over the course of 5 days, mice learned to lick within the reward zone to receive the water reward. This visual detection task was used to investigate V1 neuronal activity before, during and after learning \cite{henschke2020reward}. The goal of the experiments was to elucidate how repeated exposure to a stimulus modulates neural population responses, particularly in the presence of a stimulus-associated reward. Our analysis was based on deconvolved spike trains instead of the calcium transients. Spiking activity had been reconstructed using the MLspike algorithm \cite{deneux2016accurate} (see \cite{henschke2020reward} for more details). The data we used in this study were limited to one mouse on day 4 of the experiment when the animal was an expert at the task. Moreover, in order to provide a proof-of-concept example and illustrate the complete vine copula decomposition as well as the performance of the NSF framework, we selected a subset of 5 neurons out of the 102 V1 neurons that were monitored in total for that particular mouse. Each of the 5 selected neurons showed non-negligible positive or negative rank correlation with every other neuron in that subset (Kendall's $\tau> 0.3 $ or Kendall's $\tau < -0.3 $), suggesting that they might be part of a module warranting a more detailed investigation of the dependence structures within.

\section{Results}
\subsection{Validation on artificial data}
In order to demonstrate the potential of the NSF vine copula framework to capture multivariate dependencies of arbitrary shape and data type (continuous vs discrete) we conducted a simulation study. The set of generated samples that were used for training the NSF (n=5000) included cases with 3 different types of copulas (Clayton, Frank and Gaussian) with each dictating a different set of dependence structures for weak and stronger dependencies, continuous and discrete data as well as 4 and 8 dimensions. This ensemble provided a wide range of settings from simpler symmetric dependencies to skewed and heavy tailed interactions that can be encountered in neural population spiking responses.

Overall, performance of NSF as assessed by the KL divergence of the NSF-generated copula samples with those from the ground truth copulas was broadly within comparable levels to that of the KDE estimators while Bernstein estimators often performed the worst (Figure 2 and Figure 3). However, it is worth noting that relative performances varied slightly on a case-by-case level. For example, with weaker dependencies in 4 dimensional data with all copulas, NSF performed slightly but significantly worse than the other estimators (Kruskal-Wallis tests, $p<0.05$) except Bernstein estimators (Figure 2). The latter even outperformed NSF in one exceptional case with weakly dependent 4D discrete data with Frank copulas.  This trend of slightly worse NSF performance relative to all else except Bernstein estimators was also observed in 4D continuous and discrete data for all copulas with stronger dependencies (Figure 3). However, NSF closed that gap and performed similarly with the group of KDE estimators in cases where data was 8D, either continuous or discrete and was entangled with Clayton copulas with weak dependencies or Frank copulas with both weak and stronger dependencies (Kruskal-Wallis tests, $p>0.05$). Notably, in the case of Clayton copulas with stronger dependencies for 8D data, NSF outperformed all other estimators (Kruskal-Wallis tests, $p<0.05$). These findings might suggest that the flexibility of the NSF framework can be more beneficial with higher data dimensionality and dependence structures that are characterized by heavier tails. 

\begin{figure}[h!]
    \centering
    \includegraphics[scale=0.5]{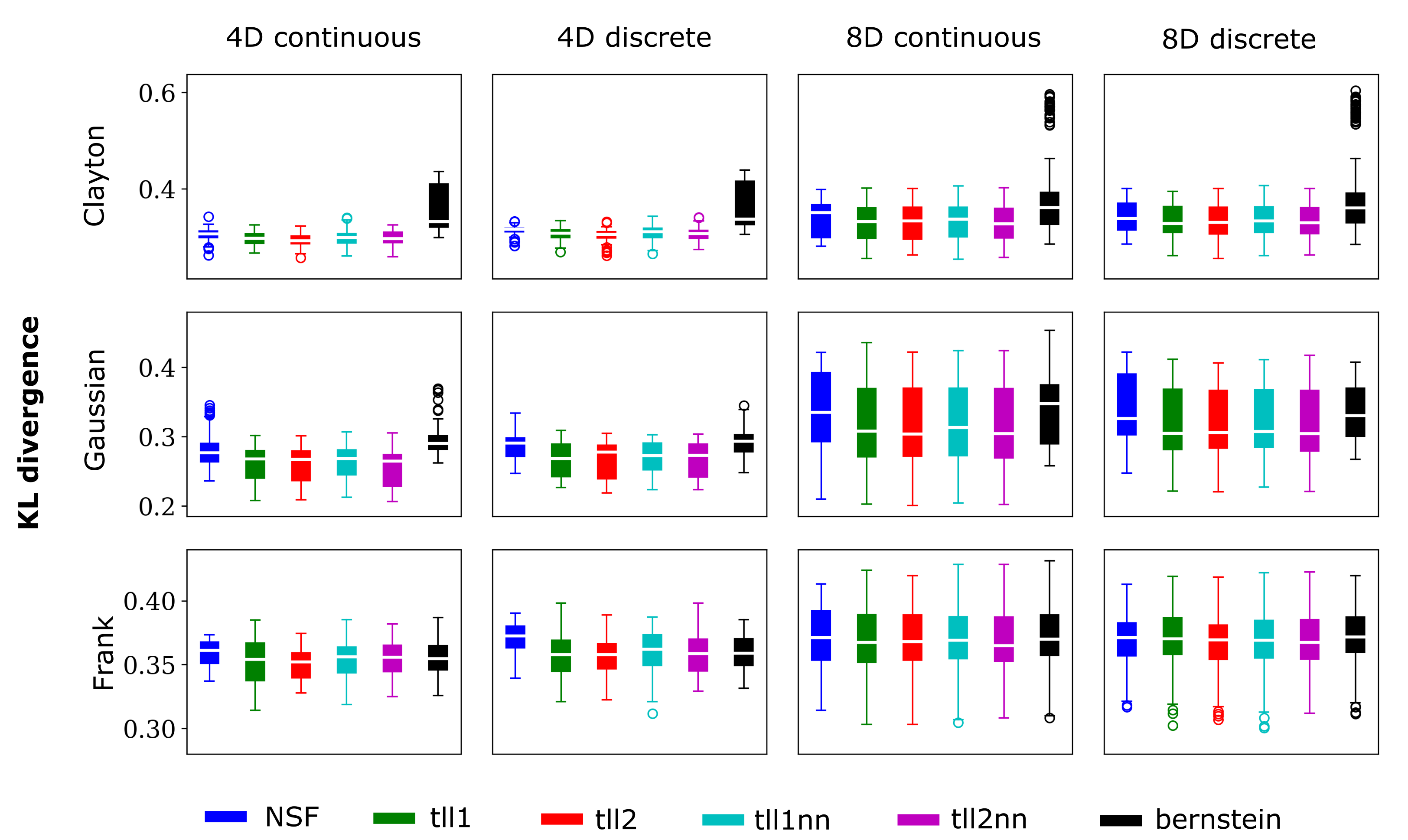}
    \caption{NSFs perform comparably to existing non-parametric estimators. Boxplots of performance of NSFs on all bivariate copulas from artificial data compared to Kernel Density Estimators with either log-linear local likelihood estimation (tll1), log-quadratic local likelihood estimation (tll2), log-linear local likelihood estimation and nearest neighbour bandwidths (tll1nn), log-quadratic local likelihood estimation and nearest neighbour bandwidths (tll2nn) and Bernstein estimator. Simulations shown in this figure had weak dependencies described by Clayton ($\theta = 2$), Gaussian ($0.4$ in the off-diagonal) and Frank copulas ($\theta = 3$) for 4 and 8 dimensional vines with continuous and discrete variables.}
    \label{fig:fig2}
\end{figure}

\begin{figure}[h!]
    \centering
    \includegraphics[scale=0.5]{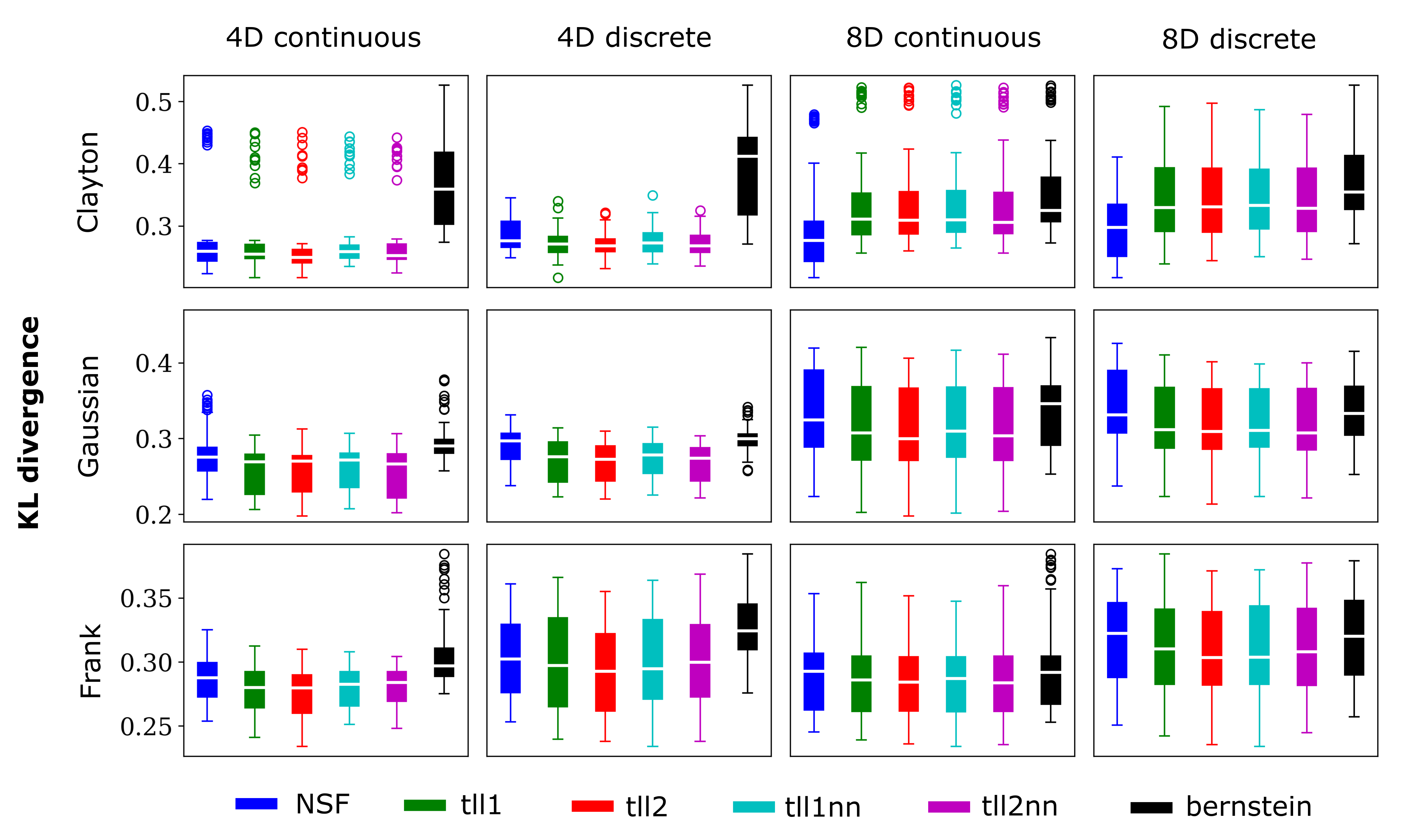}
    \caption{Same conventions as Figure 2 but for simulations with strong dependencies described by Clayton ($\theta = 5$), Gaussian ($0.8$ in the off-diagonal) and Frank copulas ($\theta = 7$) .}
    \label{fig:fig3}
\end{figure}

As copulas offer a detailed and invariant to marginal statistics view into multivariate dependencies, their negative entropy provides an accurate estimate of mutual information between variables. Thus, calculating copula entropies can be useful not only in undestanding coordinated information processing but also in BCI settings where dependencies might be an important feature needing to be accounted for. Despite the largely similar performance to KDEs, NSFs showed a remarkable advantage regarding drawing samples from the trained models and estimating copula entropies. Inverting the kernel transformation in KDE estimators and rejection sampling in Bernstein estimators are considerably more computationally expensive compared to a flexible and substantially faster sampling scheme in NSFs which directly approximate the CDF and inverse CDF of margins and copulas (Figure 4).

Plotting the copula entropies from all the estimators against the KL divergence for every particular iteration of fitting in every bivariate copula from the vine revealed an inverse relationship between the two quantities. Namely, better performance appeared to relate to higher copula entropy and thus mutual information for all distinct copulas, vine dimensions and data types (Figure 5). This could mean that bivariate copulas with higher KL divergence were overfit to the point of diminishing the informational content of the interaction captured by the copula. It is noteworthy that Clayton copulas from 8D vines that were well fit by NSFs exhibited significantly higher copula entropy compared to the other estimators (Kruskal-Wallis, $p < 0.01$). This could indicate an potential advantage of NSFs over the other estimators with heavy-tailed data, which might warrant further future investigation. 
\begin{figure}[h!]
    \centering
    \includegraphics[scale=0.4]{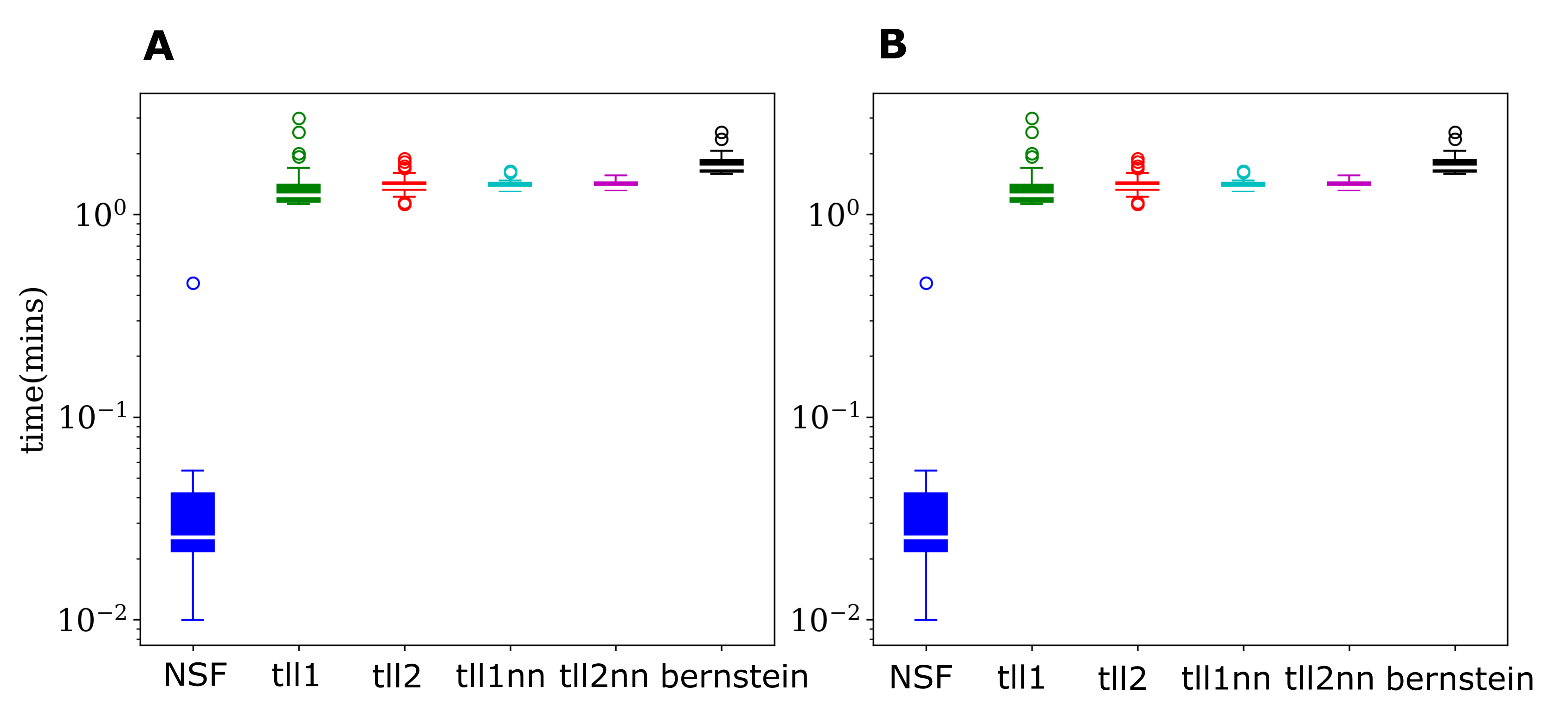}
    \caption{NSFs vastly outperform other non-parametric estimators on sampling. \textbf{A}. Boxplots of time (mins) required to sample from trained NSF compared to the other non-parametric estimators. The vertical axis is plotted on logarithmic scale. \textbf{B}. Boxplots of time (mins) required to estimate copula entropy from the trained NSF compared to the other non-parametric estimators.}
    \label{fig:fig4}
\end{figure}

\begin{figure}[h!]
    \centering
    \includegraphics[scale=0.43]{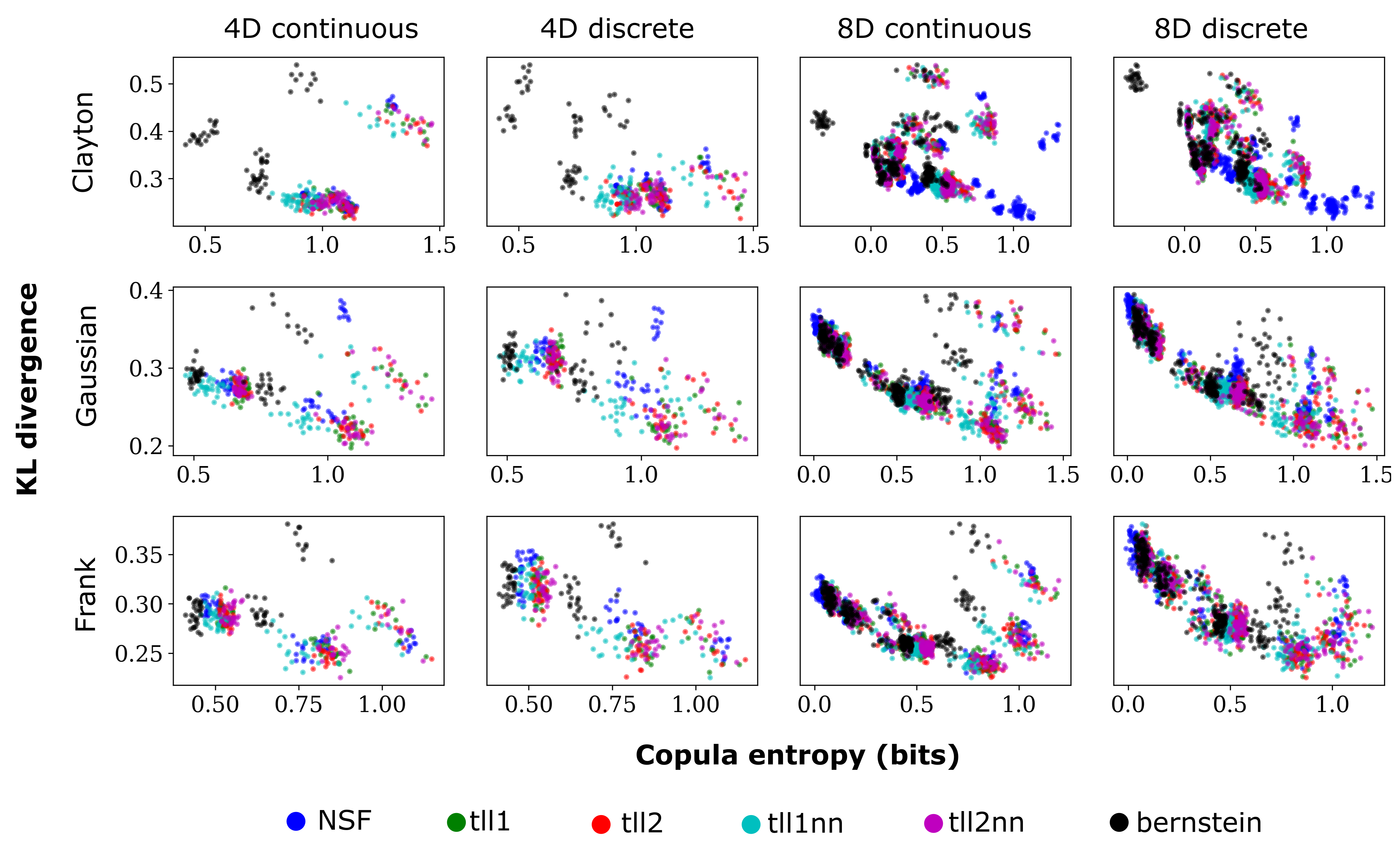}
    \caption{Scatter plots of performance against copula entropy for NSFs versus other non-parametric estimators on artificial data. All other conventions are the same as Figures 2 and 3. Simulations shown in this figure had strong dependencies.}
    \label{fig:fig5}
\end{figure}

\subsection{NSF elucidate dependence structures in rodent V1 spiking activity and behavior}

Having validated the vine flow copula framework with artificial data, we subsequently focused on spiking activity from a subset of 5 V1 layer 2/3 neurons while the animal was navigating a virtual reality corridor (Figure 6A) \cite{henschke2020reward}. A parametric vine copula framework with Gaussian processes \cite{kudryashova2022parametric} was recently applied to calcium transients from the same V1 layer 2/3 neurons \cite{henschke2020reward}. In the present study, we focused instead on modeling the deconvolved spiking activity with flow-based vine copulas. Our use of a non-parametric tool aimed at escaping potentially restricting assumptions that parametric copula families might place on the dependence structures among neurons. Our analysis aimed to detect such dependencies both as a function of time and position since previous findings indicate that V1 neuronal activity can be modulated by behaviorally relevant variables such as a reward being given at a particular location.

Raster plots across all trials on day 4 for the subset of the selected 5 neurons in Figure 6A illustrate how some of the neurons were more active for a short span of the virtual corridor within and after the reward zone (120-140 cm) while the activity of others was more spread out across the corridor and fell off at the onset of the reward zone. The strength of rank correlation between pairs of these neurons can be assessed by measuring the Kendall's $\tau$ from their spiking activities. However, this approach reduces the study of dependencies into single numbers that only provide a limited description of the interactions. In contrast, a copula-based approach can provide a detailed account of the actual shapes of neuronal dependencies which are usually characterized by a concentration of probability mass in the tail regions.

In similar fashion to the analysis on artificial data, we fit a C-vine whereby NSFs were fit to the spiking distributions of the neurons, i.e. the margins as a first step and then NSFs were sequentially fit to the empirical copulas (blue scatter plots in Figure 6B) from the surface tree level to the deeper one. These empirical copulas were obtained by transforming the data with the distributional transform. The 5 neurons were ordered according to the degree to which they exhibited statistical dependence with the other neurons as measured by the sum of Kendall's $\tau$ for each neuron. The copula densities (Figure 6C) and samples (red scatter plots in Figure 6D) from the trained NSFs were able to accurately capture the dependence structures between the neurons. All interactions were characterized by the presence of heavy tails in different corners of the uniform cube. Heavy tail dependencies at the top right part of the cube signified neurons that were more co-dependent when both neurons were active compared to other activity regimes (e.g. Figure 6C top row $1,2$ and $1,5$). For example, neurons like neuron 1 and neuron 2 displayed weak or no significant interaction until their activity was modulated by experimental or behavioral variables that are associated with the reward zone. Conversely, heavy tails at the bottom right (Figure 6C $2,3|1$) or top left (Figure 6C $4,5|1,2,3$) signified inverse relations of spiking activity between these neurons, i.e. being active at different locations in the corridor. Furthermore, it is worth noting that the probability mass concentration in the tail regions was different among neurons, with some pairs displaying lighter tails than others (e.g. Figure 6C $3,4|1,2$).

\begin{figure}[h!]
    \centering
    \includegraphics[scale=0.33]{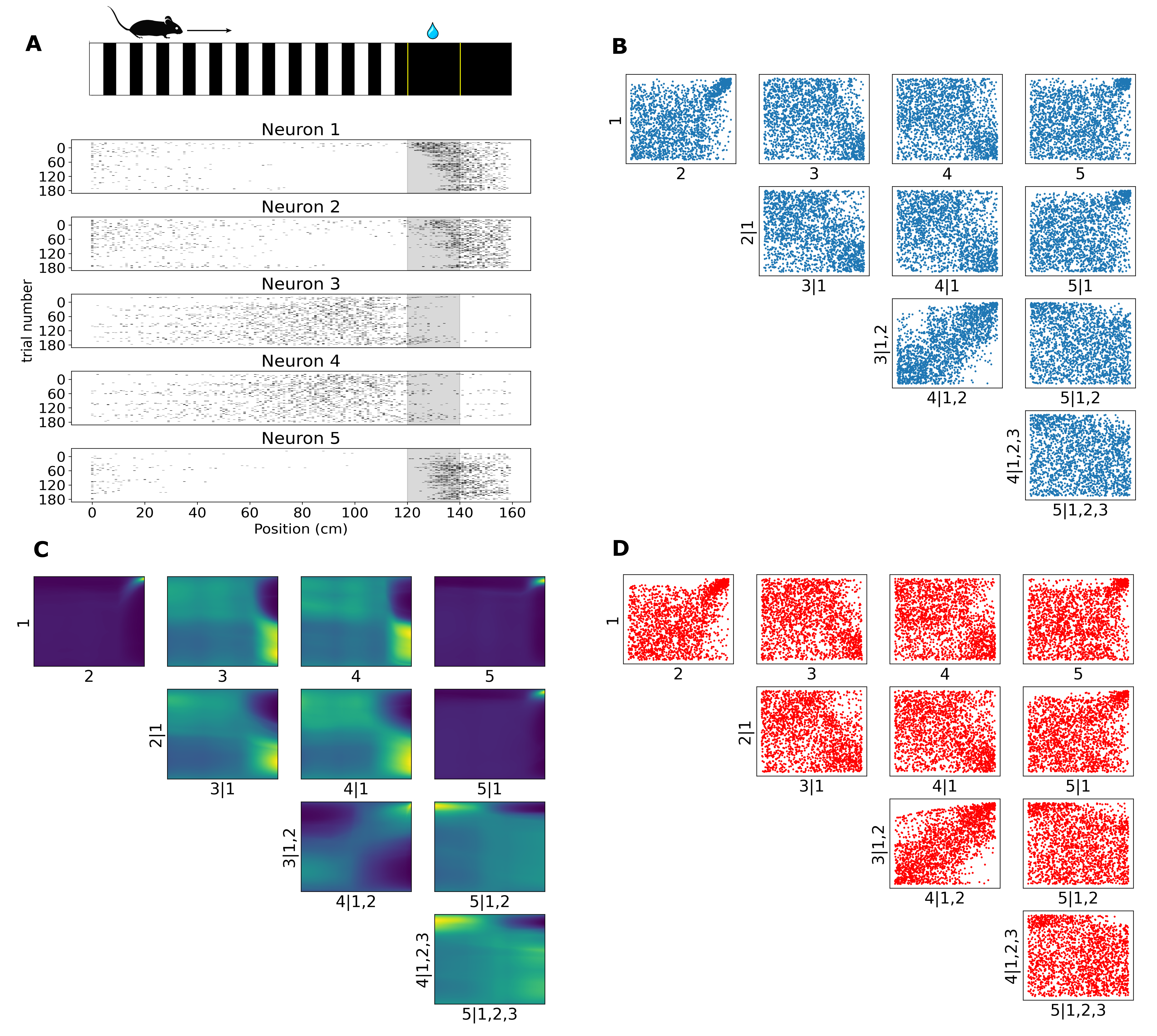}
    \caption{Tail dependencies in position dependent V1 activity are captured by NSF vine copulas. \textbf{A}. Illustration of mouse navigating a virtual environment with grating stimuli until it reaches the reward zone (120-140 cm) where it is required to lick in order to receive water reward. Raster plots for 5 neurons in V1 rodent data across trials and position (bin size of 2.5cm) in the virtual corridor. Grey region denotes the reward zone \textbf{B}. Scatter plots of empirical copula samples (blue dots) in a 5D vine extracted from the spike data. Axis labels denote which unconditional or conditional neuron margins have been transformed to copula space. Margin indices after vertical slash denote those in the conditioning set \textbf{C.} Copula densities of the 5-D vine from the trained NSF. \textbf{D.} Simulated samples (red dots) from the trained NSF copula densities of the 5-D vine. }
    \label{fig:fig6}
\end{figure}

As neuronal activity in V1 is modulated by introducing a reward in a certain location of the virtual corridor, it is also interesting to investigate neural responses and dependencies in time around that reward. Therefore, we also analyzed neural interactions as revealed by the spike counts of the same 5 neurons 3.5 seconds before and after the reward (bin size was 300~ms). Only successful trials where the mouse licked within the reward zone were included in this analysis. Raster plots in Figure 7A show a variety of spiking patterns relative to the timing of the reward across trials.  The copula densities (Figure 7C) and samples (red scatter plots in Figure 7D) from the trained NSFs closely captured the dependence structures as before. Moreover, as before, these dependence structures were characterized by heavy tailed regions either in the top right (e.g. $1,2$ in Figure 7C) or top left corners (e.g. $2,5|1$ in Figure 7C) indicating stronger neuronal interactions for some regimes of spiking activity and not otfhers. The more apparent block structure in this case compared to the previous one was a result of the fewer states of spike counts with the bin size we chose. For example, probability mass in the copula space for neurons that would fire from 0 to 5 spikes in a given time bin, would have to be assigned to blocks that correspond to the relative proportions of jointly observed spike counts. 

\begin{figure}[h!]
    \centering
    \includegraphics[scale=1.05]{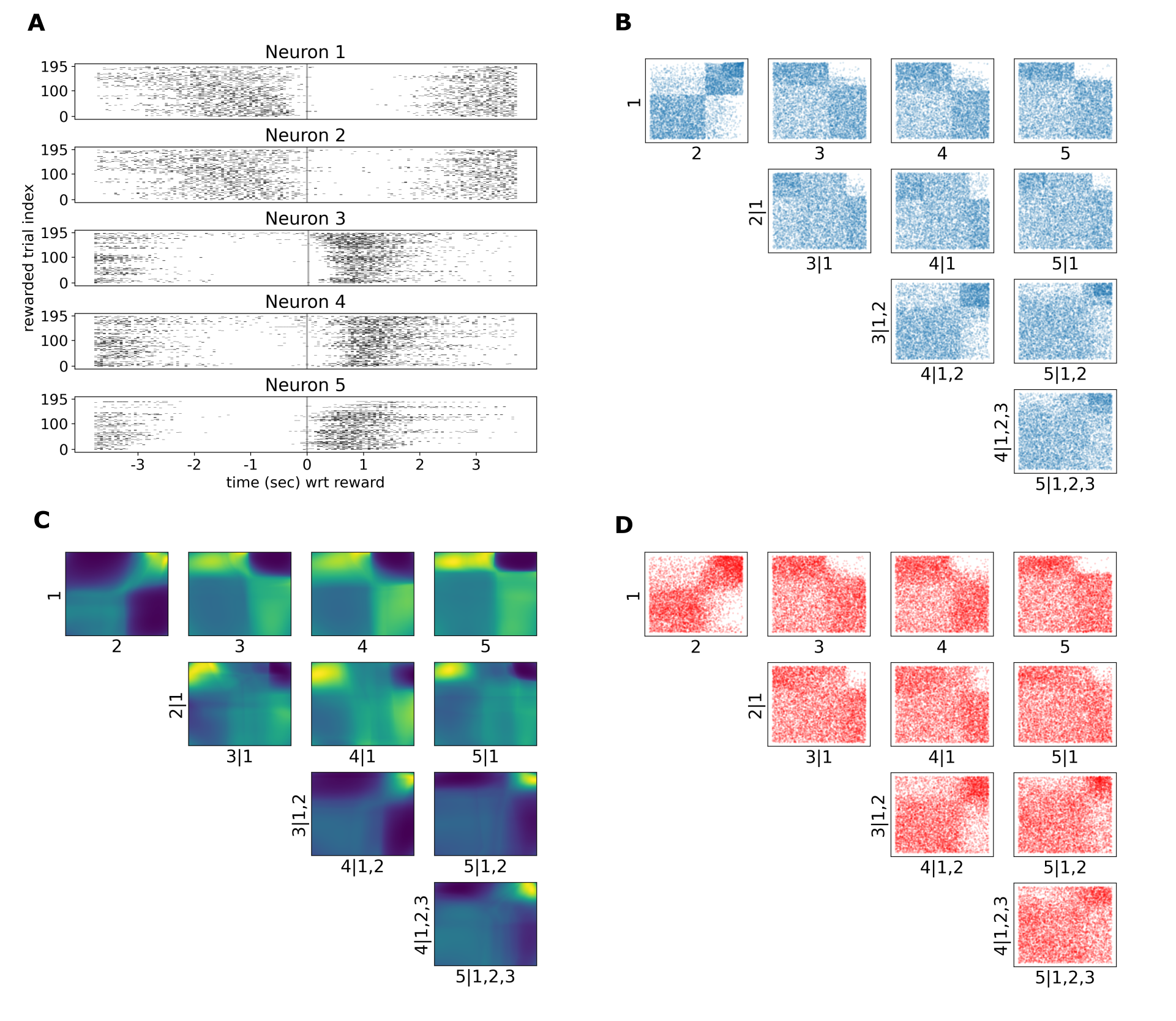}
    \caption{Tail dependencies in time dependent V1 activity are captured by NSF vine copulas. \textbf{A}. Raster plots for same 5 neurons in V1 rodent data across trials and time with respect to reward (3.5 seconds before until 3.5 seconds after reward) in the virtual corridor. Grey vertical line denotes reward acquisition. \textbf{B}. Scatter plots of empirical copula samples (blue dots) from the 5-D vine. All other conventions are the same with Figure 6 \textbf{C.} Copula densities of the 5-D vine from the trained NSF. \textbf{D.} Simulated samples (red dots) from the trained NSF copula densities of the 5-D vine. }
    \label{fig:fig7}
\end{figure}

A copula-based approach can also offer a tool for studying dependencies of neuronal activity with behavioral variables which might exhibit vastly different statistics, such as running speed and number of licks. In Figure 8 we showcase how copulas modelled with NSFs can elucidate the shape of interaction of running speed and licks with the spiking activity of an example neuron (neuron 16 from the dataset, which is not included in the 5 neurons in the previous analysis) binned with respect to position (bin size was 2.5 cm) in the virtual corridor. This neuron increased its activity considerably within the reward zone (Figure 8B) which coincided with greatly reduced running speed (Figure 8A) as the mouse stopped to lick (Figure 8C) and receive the reward. Licking was thus positively co-dependent with spiking activity, $\tau = 0.25, p < 0.001$ for number of licks and running speed was negatively co-dependent with spiking activity $\tau = -0.21, p < 0.001$. However, transforming the variables to copula space revealed mostly uniform copulas with a relatively heavier tail in the bottom right for running speed (copulas in Figure 8A) and in the top right for licking (copulas in Figure 8C). This finding indicated that these behavioral variables had a rather weak or no interaction for most of the span of the virtual corridor except for the part when the neuron was mostly active, which was the reward zone.

\begin{figure}[h!]
    \centering
    \includegraphics[scale=0.45]{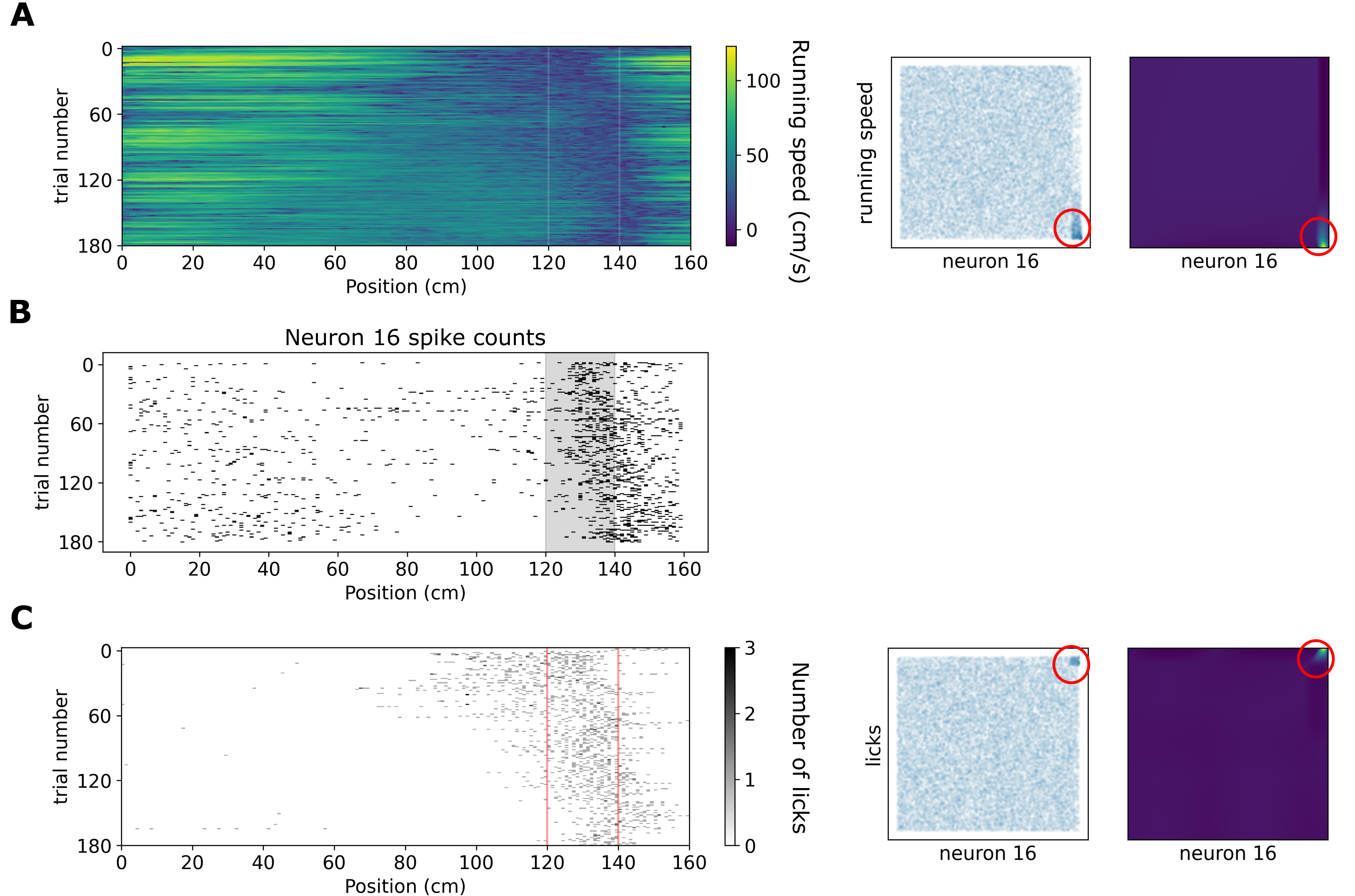}
    \caption{Dependencies of spiking activity with behavioral variables. \textbf{A}. Left: Color plot of running speed (cm/sec) across the virtual corridor for all trials. White vertical lines denote beginning and end of reward zone. Center: Empirical copula of neuron 16 spiking activity (discrete) with running speed (continuous). Right: Copula density from the trained NSF. Red circles indicate probability concentration in the tail. \textbf{B}. Raster plot of neuron 16 spiking activity across trials and position in the virtual corridor. \textbf{C.} Left: Grayscale-coded plot of number of licks across the virtual corridor for all trials. Red vertical lines denote beginning and end of reward zone. Center: Empirical copula of neuron 16 spiking activity (discrete) with number of licks (discrete). Right: Copula density from the trained NSF }
    \label{fig:fig6}
\end{figure}

Considering all the aforementioned, the flow-based vine copula framework shows remarkable promise for flexibly capturing complicated interaction patterns within neural populations. The rich picture of these interactions and potential insights into neural circuit function would have remained undetected by only measuring pairwise rank correlations or pairwise Pearson correlations which would not have indicated any heavy tailed interactions.

\section{Discussion}
In this study, we proposed a fully non-parametric approach for estimating vine copula densities for both continuous and discrete data using NSF, a subtype of normalizing flows. Overall, the framework performed comparably to existing non-parametric approaches on artificial data while benefiting from more flexible and faster sampling. Furthermore, we demonstrated how flow-based vine copulas can shed light on the structure of dependencies in neural spiking responses. 

The intricate shapes in bivariate copulas of spikes involving skewness and heavy tails would have been assumed as non-existent in a conventional approach based on pairwise linear correlations. Additionally, the arrangement of the discovered copulas into block structures (Figure 6B) would have been harder to capture by commonly applied parametric copula models (e.g. Clayton, Frank or Gaussian copulas) and thus lead to misleading conclusions. Therefore, we showed that non-parametric modeling with normalizing flows can be a valuable tool especially in the case of copula-based methods for discrete data and mixed data.

The development of such tools is crucial for understanding how coordinated neural activity transmits information about external stimuli or internal signals that guide planning and action. Copula-based methods have the capacity to provide a description of neural activity that includes the intricacies of dependencies and allows for higher-order interactions to occur within a certain set of conditions specified by the vine structure. Such descriptions can potentially offer significant insights that will aid and inform the development of neural coding theory. At the same time, decoding models that take into account the aforementioned features elucidated by copula methods, could potentially be valuable for the development of reliable Brain-Computer Interfaces. Some lines of evidence suggest that higher-order interaction patterns might have a significant presence and role in shaping collective neural dynamics and conveying information \cite{ohiorhenuan2010sparse,yu2011higher,shimazaki2012state} but further research needs to be conducted to gain a deeper understanding of the processes involved.

A limitation of our study is that the type of normalizing flows we used was designed to model variables with continuous data. We therefore included an additional rounding operation for discrete margins that were generated by trained NSF. This issue with discrete margins could be addressed in future work that can improve upon the framework by incorporating normalizing flow models that are more naturally suited to handle discrete or mixed data and do not need the ad hoc modifications that were employed in the present project. SurVAE flows that were developed recently \cite{nielsen2020survae} as an attempt to unify variational autoencoders and normalizing flows might potentially be better suited for discrete data as the kind of surjective transformations may better account for discontinuities in discrete cumulative distribution functions.

Future directions include the analysis of entire neural populations as opposed to small subsets of a population. A possible extension to modelling population dependencies with vine copula methods can also involve leveraging the power of dimensionality reduction methods. A well established trait of neural population activity is that latent collective dynamics that drive responses occupy a subspace that is much lower in dimensionality than the number of neurons recorded \cite{cunningham2014dimensionality}. Thus,
dimensionality reduction methods could be applied prior to copula modeling so that the latter can be employed on a set of low dimensional latent factors that capture the most prominent latent processes. Combining such methods can potentially provide even less computationally demanding frameworks that can be useful for clinical translation.

\section*{Conflict of Interest Statement}

The authors declare that the research was conducted in the absence of any commercial or financial relationships that could be construed as a potential conflict of interest.

\section*{Author Contributions}

LM and AO conceptualized the study and designed the methodology. LM conducted data analysis, visualization of findings under the supervision of AO, and wrote the first draft version. TA collected and curated the neuronal recordings data. LM, TA and AO reviewed and edited the article. All authors approved the submitted version.

\section*{Funding}
This work was supported by the Engineering and Physical Sciences Research Council (grant [EP/S005692/1], to AO) and the Precision Medicine Doctoral Training Programme (Medical Research Council grant number [MR/N013166/1], to TA). The funders had no role in study design, data collection and analysis, decision to publish, or preparation of the manuscript.

\section*{Acknowledgments}
We would like to thank Nathalie Rochefort and Tom Flossman for constructive feedback on the analysis and manuscript.

\bibliographystyle{unsrt}  
\bibliography{main_textMV.bbl}

\begin{thebibliography}{10}

\bibitem{gao2015simplicity}
Peiran Gao and Surya Ganguli.
\newblock On simplicity and complexity in the brave new world of large-scale
  neuroscience.
\newblock {\em Current opinion in neurobiology}, 32:148--155, 2015.

\bibitem{chen2012lotos}
Xiaowei Chen, Ulrich Leischner, Zsuzsanna Varga, Hongbo Jia, Diana Deca,
  Nathalie~L Rochefort, and Arthur Konnerth.
\newblock Lotos-based two-photon calcium imaging of dendritic spines in vivo.
\newblock {\em Nature protocols}, 7(10):1818--1829, 2012.

\bibitem{mcfarland2017eeg}
DJ~McFarland and JR~Wolpaw.
\newblock Eeg-based brain--computer interfaces.
\newblock {\em current opinion in Biomedical Engineering}, 4:194--200, 2017.

\bibitem{jun2017fully}
James~J Jun, Nicholas~A Steinmetz, Joshua~H Siegle, Daniel~J Denman, Marius
  Bauza, Brian Barbarits, Albert~K Lee, Costas~A Anastassiou, Alexandru Andrei,
  {\c{C}}a{\u{g}}atay Ayd{\i}n, et~al.
\newblock Fully integrated silicon probes for high-density recording of neural
  activity.
\newblock {\em Nature}, 551(7679):232--236, 2017.

\bibitem{chaudhary2016brain}
Ujwal Chaudhary, Niels Birbaumer, and Ander Ramos-Murguialday.
\newblock Brain--computer interfaces for communication and rehabilitation.
\newblock {\em Nature Reviews Neurology}, 12(9):513--525, 2016.

\bibitem{hurwitz2021building}
Cole Hurwitz, Nina Kudryashova, Arno Onken, and Matthias~H Hennig.
\newblock Building population models for large-scale neural recordings:
  Opportunities and pitfalls.
\newblock {\em Current opinion in neurobiology}, 70:64--73, 2021.

\bibitem{zohary1994correlated}
Ehud Zohary, Michael~N Shadlen, and William~T Newsome.
\newblock Correlated neuronal discharge rate and its implications for
  psychophysical performance.
\newblock {\em Nature}, 370(6485):140--143, 1994.

\bibitem{brown2004multiple}
Emery~N Brown, Robert~E Kass, and Partha~P Mitra.
\newblock Multiple neural spike train data analysis: state-of-the-art and
  future challenges.
\newblock {\em Nature neuroscience}, 7(5):456--461, 2004.

\bibitem{moreno2014information}
Rub{\'e}n Moreno-Bote, Jeffrey Beck, Ingmar Kanitscheider, Xaq Pitkow, Peter
  Latham, and Alexandre Pouget.
\newblock Information-limiting correlations.
\newblock {\em Nature neuroscience}, 17(10):1410--1417, 2014.

\bibitem{kohn2016correlations}
Adam Kohn, Ruben Coen-Cagli, Ingmar Kanitscheider, and Alexandre Pouget.
\newblock Correlations and neuronal population information.
\newblock {\em Annual review of neuroscience}, 39:237--256, 2016.

\bibitem{averbeck2006neural}
Bruno~B Averbeck, Peter~E Latham, and Alexandre Pouget.
\newblock Neural correlations, population coding and computation.
\newblock {\em Nature reviews neuroscience}, 7(5):358--366, 2006.

\bibitem{ecker2010decorrelated}
Alexander~S Ecker, Philipp Berens, Georgios~A Keliris, Matthias Bethge, Nikos~K
  Logothetis, and Andreas~S Tolias.
\newblock Decorrelated neuronal firing in cortical microcircuits.
\newblock {\em science}, 327(5965):584--587, 2010.

\bibitem{onken2009analyzing}
Arno Onken, Steffen Gr{\"u}new{\"a}lder, Matthias~HJ Munk, and Klaus Obermayer.
\newblock Analyzing short-term noise dependencies of spike-counts in macaque
  prefrontal cortex using copulas and the flashlight transformation.
\newblock {\em PLoS computational biology}, 5(11):e1000577, 2009.

\bibitem{kudryashova2022parametric}
Nina Kudryashova, Theoklitos Amvrosiadis, Nathalie Dupuy, Nathalie Rochefort,
  and Arno Onken.
\newblock Parametric copula-gp model for analyzing multidimensional neuronal
  and behavioral relationships.
\newblock {\em PLoS computational biology}, 18(1):e1009799, 2022.

\bibitem{pillow2008spatio}
Jonathan~W Pillow, Jonathon Shlens, Liam Paninski, Alexander Sher, Alan~M
  Litke, EJ~Chichilnisky, and Eero~P Simoncelli.
\newblock Spatio-temporal correlations and visual signalling in a complete
  neuronal population.
\newblock {\em Nature}, 454(7207):995--999, 2008.

\bibitem{ohiorhenuan2010sparse}
Ifije~E Ohiorhenuan, Ferenc Mechler, Keith~P Purpura, Anita~M Schmid, Qin Hu,
  and Jonathan~D Victor.
\newblock Sparse coding and high-order correlations in fine-scale cortical
  networks.
\newblock {\em Nature}, 466(7306):617--621, 2010.

\bibitem{yu2011higher}
Shan Yu, Hongdian Yang, Hiroyuki Nakahara, Gustavo~S Santos, Danko Nikoli{\'c},
  and Dietmar Plenz.
\newblock Higher-order interactions characterized in cortical activity.
\newblock {\em Journal of neuroscience}, 31(48):17514--17526, 2011.

\bibitem{shimazaki2012state}
Hideaki Shimazaki, Shun-ichi Amari, Emery~N Brown, and Sonja Gr{\"u}n.
\newblock State-space analysis of time-varying higher-order spike correlation
  for multiple neural spike train data.
\newblock {\em PLoS computational biology}, 8(3):e1002385, 2012.

\bibitem{montangie2017higher}
Lisandro Montangie and Fernando Montani.
\newblock Higher-order correlations in common input shapes the output spiking
  activity of a neural population.
\newblock {\em Physica A: Statistical Mechanics and its Applications},
  471:845--861, 2017.

\bibitem{michel2006costs}
Melchi~M Michel and Robert~A Jacobs.
\newblock The costs of ignoring high-order correlations in populations of model
  neurons.
\newblock {\em Neural computation}, 18(3):660--682, 2006.

\bibitem{jaworski2012copulae}
Piotr Jaworski, Fabrizio Durante, and Wolfgang~Karl H{\"a}rdle.
\newblock Copulae in mathematical and quantitative finance.
\newblock In {\em Proceedings of the workshop held in Cracow}, volume~10,
  page~11. Springer, 2012.

\bibitem{faugeras2017inference}
Olivier~P Faugeras.
\newblock Inference for copula modeling of discrete data: a cautionary tale and
  some facts.
\newblock {\em Dependence Modeling}, 5(1):121--132, 2017.

\bibitem{genest2007primer}
Christian Genest and Johanna Ne{\v{s}}lehov{\'a}.
\newblock A primer on copulas for count data.
\newblock {\em ASTIN Bulletin: The Journal of the IAA}, 37(2):475--515, 2007.

\bibitem{song2009joint}
Peter X-K Song, Mingyao Li, and Ying Yuan.
\newblock Joint regression analysis of correlated data using gaussian copulas.
\newblock {\em Biometrics}, 65(1):60--68, 2009.

\bibitem{de2011copula}
Alex~R de~Leon and Bohsiu Wu.
\newblock Copula-based regression models for a bivariate mixed discrete and
  continuous outcome.
\newblock {\em Statistics in Medicine}, 30(2):175--185, 2011.

\bibitem{smith2012estimation}
Michael~S Smith and Mohamad~A Khaled.
\newblock Estimation of copula models with discrete margins via bayesian data
  augmentation.
\newblock {\em Journal of the American Statistical Association},
  107(497):290--303, 2012.

\bibitem{panagiotelis2012pair}
Anastasios Panagiotelis, Claudia Czado, and Harry Joe.
\newblock Pair copula constructions for multivariate discrete data.
\newblock {\em Journal of the American Statistical Association},
  107(499):1063--1072, 2012.

\bibitem{onken2016mixed}
Arno Onken and Stefano Panzeri.
\newblock Mixed vine copulas as joint models of spike counts and local field
  potentials.
\newblock {\em Advances in Neural Information Processing Systems}, 29, 2016.

\bibitem{aas2009pair}
Kjersti Aas, Claudia Czado, Arnoldo Frigessi, and Henrik Bakken.
\newblock Pair-copula constructions of multiple dependence.
\newblock {\em Insurance: Mathematics and economics}, 44(2):182--198, 2009.

\bibitem{racine2015mixed}
Jeffrey~S Racine.
\newblock Mixed data kernel copulas.
\newblock {\em Empirical Economics}, 48(1):37--59, 2015.

\bibitem{geenens2017probit}
Gery Geenens, Arthur Charpentier, and Davy Paindaveine.
\newblock Probit transformation for nonparametric kernel estimation of the
  copula density.
\newblock {\em Bernoulli}, 23(3):1848--1873, 2017.

\bibitem{schallhorn2017d}
Niklas Schallhorn, Daniel Kraus, Thomas Nagler, and Claudia Czado.
\newblock D-vine quantile regression with discrete variables.
\newblock {\em arXiv preprint arXiv:1705.08310}, 2017.

\bibitem{nagler2017nonparametric}
Thomas Nagler, Christian Schellhase, and Claudia Czado.
\newblock Nonparametric estimation of simplified vine copula models: comparison
  of methods.
\newblock {\em Dependence Modeling}, 5(1):99--120, 2017.

\bibitem{rezende2015variational}
Danilo Rezende and Shakir Mohamed.
\newblock Variational inference with normalizing flows.
\newblock In {\em International conference on machine learning}, pages
  1530--1538. PMLR, 2015.

\bibitem{papamakarios2021normalizing}
George Papamakarios, Eric Nalisnick, Danilo~Jimenez Rezende, Shakir Mohamed,
  and Balaji Lakshminarayanan.
\newblock Normalizing flows for probabilistic modeling and inference.
\newblock {\em Journal of Machine Learning Research}, 22(57):1--64, 2021.

\bibitem{wiese2019copula}
Magnus Wiese, Robert Knobloch, and Ralf Korn.
\newblock Copula \& marginal flows: Disentangling the marginal from its joint.
\newblock {\em arXiv preprint arXiv:1907.03361}, 2019.

\bibitem{kamthe2021copula}
Sanket Kamthe, Samuel Assefa, and Marc Deisenroth.
\newblock Copula flows for synthetic data generation.
\newblock {\em arXiv preprint arXiv:2101.00598}, 2021.

\bibitem{widge2014pre}
Alik~S Widge and Chet~T Moritz.
\newblock Pre-frontal control of closed-loop limbic neurostimulation by rodents
  using a brain--computer interface.
\newblock {\em Journal of neural engineering}, 11(2):024001, 2014.

\bibitem{bridges2018rodent}
Nathaniel~R Bridges, Michael Meyers, Jonathan Garcia, Patricia~A Shewokis, and
  Karen~A Moxon.
\newblock A rodent brain-machine interface paradigm to study the impact of
  paraplegia on bmi performance.
\newblock {\em Journal of neuroscience methods}, 306:103--114, 2018.

\bibitem{sklar1959fonctions}
M~Sklar.
\newblock Fonctions de repartition an dimensions et leurs marges.
\newblock {\em Publ. inst. statist. univ. Paris}, 8:229--231, 1959.

\bibitem{bedford2002vines}
Tim Bedford and Roger~M Cooke.
\newblock Vines--a new graphical model for dependent random variables.
\newblock {\em The Annals of Statistics}, 30(4):1031--1068, 2002.

\bibitem{durkan2019neural}
Conor Durkan, Artur Bekasov, Iain Murray, and George Papamakarios.
\newblock Neural spline flows.
\newblock {\em Advances in neural information processing systems}, 32, 2019.

\bibitem{czado2019analyzing}
Claudia Czado.
\newblock Analyzing dependent data with vine copulas.
\newblock {\em Lecture Notes in Statistics, Springer}, 2019.

\bibitem{bergstra2012random}
James Bergstra and Yoshua Bengio.
\newblock Random search for hyper-parameter optimization.
\newblock {\em Journal of machine learning research}, 13(2), 2012.

\bibitem{sancetta2004bernstein}
Alessio Sancetta and Stephen Satchell.
\newblock The bernstein copula and its applications to modeling and
  approximations of multivariate distributions.
\newblock {\em Econometric theory}, 20(3):535--562, 2004.

\bibitem{nelsen2007introduction}
Roger~B Nelsen.
\newblock {\em An introduction to copulas}.
\newblock Springer Science \& Business Media, 2007.

\bibitem{wang2009divergence}
Qing Wang, Sanjeev~R Kulkarni, and Sergio Verd{\'u}.
\newblock Divergence estimation for multidimensional densities via $ k
  $-nearest-neighbor distances.
\newblock {\em IEEE Transactions on Information Theory}, 55(5):2392--2405,
  2009.

\bibitem{nkl_hartland}
Nathan Hartland.
\newblock Kl divergence estimators, 2021.

\bibitem{jenison2004shape}
Rick~L Jenison and Richard~A Reale.
\newblock The shape of neural dependence.
\newblock {\em Neural computation}, 16(4):665--672, 2004.

\bibitem{henschke2020reward}
Julia~U Henschke, Evelyn Dylda, Danai Katsanevaki, Nathalie Dupuy, Stephen~P
  Currie, Theoklitos Amvrosiadis, Janelle~MP Pakan, and Nathalie~L Rochefort.
\newblock Reward association enhances stimulus-specific representations in
  primary visual cortex.
\newblock {\em Current Biology}, 30(10):1866--1880, 2020.

\bibitem{deneux2016accurate}
Thomas Deneux, Attila Kaszas, Gergely Szalay, Gergely Katona, Tam{\'a}s Lakner,
  Amiram Grinvald, Bal{\'a}zs R{\'o}zsa, and Ivo Vanzetta.
\newblock Accurate spike estimation from noisy calcium signals for ultrafast
  three-dimensional imaging of large neuronal populations in vivo.
\newblock {\em Nature communications}, 7(1):1--17, 2016.

\bibitem{nielsen2020survae}
Didrik Nielsen, Priyank Jaini, Emiel Hoogeboom, Ole Winther, and Max Welling.
\newblock Survae flows: Surjections to bridge the gap between vaes and flows.
\newblock {\em Advances in Neural Information Processing Systems},
  33:12685--12696, 2020.

\bibitem{cunningham2014dimensionality}
John~P Cunningham and M~Yu Byron.
\newblock Dimensionality reduction for large-scale neural recordings.
\newblock {\em Nature neuroscience}, 17(11):1500--1509, 2014.

\end{thebibliography}

\end{document}